\documentclass[11pt,prd,a4paper,preprintnumbers,amsmath,amssymb,nofootinbib]{article}
\pdfoutput=1
\usepackage{feynmp-auto,expdlist}
\usepackage{amsmath,amsfonts,amssymb}
\usepackage{graphicx}
\usepackage{enumerate}
\usepackage{hyperref}
\usepackage{latexsym}
\usepackage{hepnicenames}
\usepackage{enumerate}
\usepackage{soul}
\usepackage[normalem]{ulem}
\usepackage{wasysym}
\usepackage{makecell}
\usepackage{bbm}

\font\tenrsfs=rsfs10 at 12pt
\font\sevenrsfs=rsfs7
\font\fiversfs=rsfs5
\newfam\rsfsfam
\textfont\rsfsfam=\tenrsfs
\scriptfont\rsfsfam=\sevenrsfs
\scriptscriptfont\rsfsfam=\fiversfs

\numberwithin{equation}{section}

\usepackage{mathrsfs}
\usepackage{braket}
\usepackage{titling}
\usepackage{amsmath}
\usepackage{slashed}
\usepackage{amssymb}
\usepackage{epsfig}
\usepackage{graphicx}
\usepackage{color}
\usepackage{rotating}
\usepackage{hyperref}
\usepackage[margin=1.in]{geometry}
\usepackage[table,xcdraw,dvipsnames]{xcolor}
\usepackage[compress,numbers,sort]{natbib}
\usepackage{colortbl}
\usepackage{pdflscape}
\usepackage{color}
\usepackage{mathtools}
\usepackage{colortbl}
\usepackage{comment}
\definecolor{Gray}{gray}{0.95}
\definecolor{RGray}{gray}{0.85}
\definecolor{CGray}{gray}{0.93}

\newcommand{\B}{{\cal V}}
\newcommand{\D}{{\cal D}}

\newcommand{\SU}{{\rm SU}}

\newcommand{\U}{{\rm U}}

\newcommand{\N}{{\cal N}}

\definecolor{nicered}{rgb}{0.7,0.1,0.1}
\definecolor{nicegreen}{rgb}{0.1,0.5,0.1}
\definecolor{red}{rgb}{1.0, 0, 0}
\definecolor{niceblue}{rgb}{0,0,0.8}
\definecolor{red}{rgb}{1.0, 0, 0}
\hypersetup{colorlinks,bookmarksopen,bookmarksnumbered,
linkcolor=blus,pdfstartview=FitH,urlcolor=rossos,citecolor=verde}
\allowdisplaybreaks

\definecolor{rosso}{cmyk}{0,1,1,0.4}
\definecolor{rossos}{cmyk}{0,1,1,0.55}
\definecolor{rossoc}{cmyk}{0,1,1,0.2}
\definecolor{blu}{cmyk}{1,1,0,0.3}
\definecolor{blus}{cmyk}{1,1,0,0.6}
\definecolor{bluc}{cmyk}{1,1,0,0.1}
\definecolor{verde}{cmyk}{0.92,0,0.59,0.25}
\definecolor{verdec}{cmyk}{0.92,0,0.59,0.15}
\definecolor{verdes}{cmyk}{0.92,0,0.59,0.4}

\def\eq#1{{Eq.~(\ref{#1})}}

\def\fig#1{{Fig.~\ref{#1}}}

\def\Table#1{{Table~\ref{#1}}}

\def\sect#1{{Sect.~\ref{#1}}}

\def\app#1{{App.~\ref{#1}}}
\def\apps#1#2{{Apps.~\ref{#1}--\ref{#2}}}
\def\Im{\mbox{Im}\,}
\def\Re{\mbox{Re}\,}

\renewcommand{\bar}{\overline}

\newcommand{\beq}{\begin{equation}}
\newcommand{\eeq}{\end{equation}}
\newcommand{\bea}{\begin{eqnarray}}
\newcommand{\eea}{\end{eqnarray}}

\renewcommand{\[}{\left[}
\renewcommand{\]}{\right]}
\renewcommand{\(}{\left(}
\renewcommand{\)}{\right)}

\begin{document}

%\thispagestyle{empty}

%\begin{flushleft}
%\hskip 9cm TTP21-014, P3H-21-037, ZU-TH 24/21
%\end{flushleft}

\begin{center}  
%\vspace{-0.cm}
{\LARGE\huge\bf\color{blus} 
What is the scale of new physics \\[2mm]
behind the muon $g-2$ ?} \\
\vspace{0.8cm}

{\bf Lukas Allwicher$^{a}$, Luca Di Luzio$^{b,c}$, Marco Fedele$^{d}$, \\ 
Federico Mescia$^{e}$, Marco Nardecchia$^{f}$}\\[7mm]

{\it $^a$Physik-Institut, Universit\"at Z\"urich, CH-8057 Z\"urich, Switzerland}\\[1mm]
{\it $^b$Dipartimento di Fisica e Astronomia `G.~Galilei', Universit\`a di Padova, Italy}\\[1mm]
{\it $^c$Istituto Nazionale Fisica Nucleare, Sezione di Padova, Italy}\\[1mm]
{\it $^d$Institut f\"ur Theoretische Teilchenphysik, Karlsruhe Institute of Technology, \\ 
D-76131 Karlsruhe, Germany}\\[1mm]
{\it $^e$Departament de F\'isica Qu\`antica i Astrof\'isica, Institut de Ci\`encies del Cosmos (ICCUB),
Universitat de Barcelona, Mart\'i i Franqu\`es 1, E-08028 Barcelona, Spain}\\[1mm]
{\it $^f$Physics Department and INFN Sezione di Roma La Sapienza, \\ 
Piazzale Aldo Moro 5, 00185 Roma, Italy}\\[1mm]

\vspace{0.3cm}
\begin{quote}

We study the constraints imposed by 
perturbative unitarity on the new physics interpretation of the muon 
$g-2$ anomaly.
%Assuming a short-distance 
%origin of the latter, 
Within a Standard Model Effective Field Theory 
(SMEFT) 
approach, 
we find that scattering amplitudes sourced by 
effective operators 
saturate perturbative unitarity at about 1 PeV.  
%around 
%900 TeV (600 TeV) for a dipole operator oriented in the U(1)$_Y$  (SU(2)$_L$) direction. 
This 
corresponds to
%is
the highest energy scale that needs to be probed 
in order to resolve the new physics origin of the muon $g-2$ anomaly. 
On the other hand, simplified models  (e.g.~scalar-fermion Yukawa theories) in which renormalizable couplings are pushed to the boundary of perturbativity still imply new on-shell states below 200 TeV. 
We finally suggest that the highest new physics scale responsible for the anomalous effect can be reached in non-renormalizable models at the PeV scale.

%unitarity bounds in the SMEFT 
%can be saturated by a strongly-coupled dynamics at the PeV scale,   
%whose origin would remain secluded to next generation high-energy particle colliders. 
\end{quote}
\thispagestyle{empty}
\end{center}

\bigskip
\tableofcontents

%\clearpage

\section{Introduction}
\label{sec:intro}

The recent measurement of the 
muon
anomalous 
magnetic moment,  
$a_\mu \equiv (g_\mu -2) / 2$, 
by the 
E989 experiment at Fermilab \cite{Abi:2021gix}, in agreement with the previous 
BNL E821 result \cite{Bennett:2006fi}, implies a 4.2$\sigma$ discrepancy from the 
Standard Model (SM) 
\beq 
\label{eq:Deltaamu} 
\Delta a_\mu \equiv a_\mu(\text{Exp}) - a_\mu(\text{SM}) = ( 251 \pm 59 ) \times 10^{-11} \, ,
\eeq
following the Muon $g-2$ Theory Initiative recommended value 
for the SM theory prediction \cite{Aoyama:2020ynm}. 
Although a recent lattice determination of the 
SM hadron vacuum polarization contribution to 
$a_\mu$ claims no sizeable deviation from the SM \cite{Borsanyi:2020mff}, 
we will 
work here under 
the hypothesis that $\Delta a_\mu$ is due to new physics. 
In particular, we will focus on the case in which new physics states are so heavy that 
their effects can be parameterized 
via the so-called SM Effective Field Theory (SMEFT) and ask the following question: 
\emph{What is the scale of new physics behind $\Delta a_\mu$?} 

This question is of practical relevance, given the futuristic possibility of resolving the new physics origin of 
$\Delta a_\mu$ via direct searches at high-energy particle colliders. 
As explored recently in 
\cite{Capdevilla:2020qel,Buttazzo:2020eyl,Capdevilla:2021rwo}, 
a muon collider seems to be the best option for this goal. 
However, while 
the very existence of the SMEFT operators contributing to $\Delta a_\mu$ 
could be tested via processes like 
$\mu^+ \mu^- \to Z(\gamma) h$ or $\mu^+ \mu^- \to t \bar t$ at a multi-TeV-scale muon collider \cite{Buttazzo:2020eyl}, 
it is less clear whether the origin of the muon $g-2$ SMEFT operators can be resolved 
via the direct production of new on-shell states 
responsible for $\Delta a_\mu$. 
This is the question that we want to address in the present work, using the 
tools of perturbative unitarity. Unitarity bounds on the new physics interpretation of $\Delta a_\mu$  
were previously considered in \cite{Capdevilla:2020qel,Capdevilla:2021rwo} focusing however 
on a specific class of renormalizable models. Here, 
we will consider instead the most conservative case in which unitarity limits 
are obtained within the SMEFT 
and 
reach a more pessimistic conclusion about the possibility of 
establishing a no-lose theorem for testing the origin of $\Delta a_\mu$ 
at a future high-energy particle collider. 

Generally speaking, 
given a low-energy determination of an EFT coefficient,  
unitarity methods 
can be used either within an EFT approach,  
in order to infer an upper bound on the scale of new physics 
unitarizing EFT scattering amplitudes,  
or within explicit new physics (renormalizable) models. 
In the latter case, one obtains a perturbativity bound on certain renormalizable couplings 
that can be translated into an upper bound on the mass of new on-shell degrees of freedom. 
In the present work we will be interested in both these approaches.  
First, 
we will consider a SMEFT analysis in 
which 
$\Delta a_\mu$
is explained in terms of 
a set of Wilson coefficients normalized to some 
cut-off scale$^2$, $C_i / \Lambda^2$, and later deal with renormalizable models 
featuring new heavy mediators that can be matched onto the SMEFT. 
Schematically, 
\beq 
\label{eq:Damustrategy}
\Delta a _\mu \sim \frac{C_i}{\Lambda^2} = \frac{\text{(loops)} \times \text{(couplings)}}{M^2_{\rm on-shell}} \, , 
\eeq 
where $M_{\rm on-shell}$ denote the mass of new on-shell states 
and we included possible loop factors in the matching between the new physics model and the SMEFT operators. 
Hence, by fixing the value of the SMEFT coefficients $C_i / \Lambda^2$ in terms of $\Delta a_{\mu}$, 
we will 
consider high-energy scatterings sourced by the associated effective operators,
determine the $\sqrt{s}$ that saturates perturbative unitarity 
(according to a standard 
criterium to be specified in \sect{sec:unitarity}) and interpret the latter as an 
upper bound on the 
scale of new physics responsible for the muon $g-2$ anomaly. Analogously, in the case of 
new physics models, we will use the unitarity tool in order to set 
perturbativity bounds on the new physics couplings and in turn (given \eq{eq:Damustrategy})
an upper limit 
on $M_{\rm on-shell}$. 
While the first approach is model-independent (barring possible degeneracies in the choice of the effective operators) 
and yields the most conservative bound on the scale of new physics, 
the second approach relies on further assumptions, but it directly connects 
to new on-shell degrees of freedom which are the prime targets of direct searches 
at high-energy particle colliders. 

The paper is structured as follows. We start in \sect{sec:unitarity} with a brief review 
of partial wave unitarity, in order to set notations and clarify the physical interpretation 
of unitarity bounds. Next, we consider unitarity bounds within a SMEFT approach (\sect{sec:SMEFT}) 
and within 
renormalizable models matching onto the SMEFT operators (\sect{sec:simplmodels}). 
Finally, we comment in \sect{sec:strongmodels} on non-renormalizable realizations which 
can saturate the 
unitarity bounds obtained in the SMEFT.   
Our main findings and implications for the direct resolution of 
the muon $g-2$ anomaly at high-energy particle colliders 
are summarized in the conclusions (\sect{sec:conclusions}). 
Technical aspects of partial wave unitarity calculations, both in the SMEFT and in renormalizable setups, 
are deferred to \apps{app:unitaritySMEFT}{app:unitarityweakly}.

\section{Partial wave unitarity}
\label{sec:unitarity}

We start with an instant review of 
partial wave unitarity, which will serve to set notations and discuss the physical 
significance of unitarity bounds. 

The key point of our analysis is the study of scattering amplitudes  with fixed total angular momentum $J$, the so-called partial waves.
Here we focus 
only on the case of $2 \to 2$ partial waves (while the $2\to 3$ scattering is discussed in \app{app:2to3scatterings}) defined as
\beq
	a^J_{fi} = 
	\frac{1}{32\pi} \int_{-1}^1 \mathrm{d} \cos\theta \, d^J_{\mu_i\mu_f}(\theta) \, \mathcal{T}_{fi}(\sqrt{s},\cos\theta) \,,
	\label{eq:partialwaves}
\eeq
with $\theta$ the scattering angle in the centre-of-mass frame,
$(2\pi)^4\delta^{(4)}(P_i-P_f)i\mathcal{T}_{fi}(\sqrt{s},\cos\theta) = \bra{f} S-1 \ket{i}$ and $S$ the $S$-matrix. Here, $d^J_{\mu_i\mu_f}$ is Wigner's 
$d$-function that arises in the construction of the two-particle incoming (outcoming) state of helicities  $\mu_{i} $ ($\mu_{f} $) onto angular momentum $J$ 
\cite{Jacob:1959at}. 
The $S$-matrix unitarity condition $S^\dagger S =1$ then yields the relation
\beq
	\frac{1}{2i}(a^J_{fi} - a^{J^*}_{if}) = \sum_h a^{J^*}_{hf} a^J_{hi} \qquad \Longrightarrow \qquad \Im(a_{ii}^J) 
	= \sum_h | a^J_{hi} |^2 \geq | a^J_{ii} |^2 \,,
	\label{eq:Imaii}
\eeq
where we have restricted ourselves to the elastic channel $h=i=f$. 
The equation on the right hand side of \eqref{eq:Imaii} defines a circle in the complex plane inside which the amplitude must lie at all orders,
\beq
	\Big(\Re a^J_{ii} \Big)^2 + \Big(\Im a^J_{ii} - \frac{1}{2} \Big)^2 \leq \frac{1}{4} \,,
\eeq
suggesting the following bound, under the assumption of real tree-level amplitudes:
\beq
	|\Re a^{J}_{ii}| \leq \frac{1}{2} \,. 
	\label{eq:partialwavebound}
\eeq 
Hence, 
in order to extract the bound, one needs to fully diagonalize the matrix $a^J$. Once this is achieved, every eigenvalue 
will give an independent constraint. 
In the presence of multiple scattering channels, it follows from 
\eq{eq:partialwavebound} that the strongest bound arises from the largest eigenvalue of $a^J$.
When the latter bound is saturated, 
it basically means that one needs a correction of at least $40$\% from 
higher orders to get back inside the unitarity circle, 
thus signaling the breakdown of perturbation theory (see e.g.~\cite{DiLuzio:2016sur,DiLuzio:2017chi}). 
Here, $a^{J}$ stands for 
the leading order expansion of the partial wave, 
both in the coupling constants 
and in external momenta over cut-off scale 
for the case of an EFT. 

Although the criterium is somewhat arbitrary, 
and hence \eq{eq:partialwavebound} should not be understood as a strict bound, 
we stick to that for historical reasons \cite{Lee:1977eg}. 
Strictly speaking, a violation of the perturbative unitarity criterium 
in \eq{eq:partialwavebound} 
should be conservatively interpreted as the onset of a regime of incalculability 
due to the breakdown of the perturbative expansion either in couplings or external momenta. 
More specifically, 
in the case of an EFT 
(where scattering amplitudes grow with energy)
the scale of unitarity violation,  
hereafter denoted as 
\beq 
\label{eq:LambdaU}
\sqrt{s} = \Lambda_U \qquad \Longrightarrow \qquad 
|\Re a^{J}_{ii}| = \frac{1}{2} \, , 
\eeq
can be associated with the onset of ``new physics'', 
where on-shell new degrees of freedom should 
manifest themselves and be kinematically accessible. 
Although one can conceive exotic UV 
completions where this is not the case \cite{Dvali:2010jz}, 
well-known 
physical systems 
behave in this way.\footnote{Most notably, $\pi\pi$ scattering in chiral perturbation theory 
yields $\Lambda_U = \sqrt{8\pi} f_\pi \simeq 460$ MeV which is not far from 
the mass of the $\rho$ meson resonance.}  
Unitarity methods can be employed in renormalizable setups as well. In this case, 
the unitarity limit corresponds to the failure 
of the 
coupling expansion 
and hence the bound on the renormalizable coupling 
can be understood as a perturbativity constraint.

\section{SMEFT} 
%approach to $\Delta a_\mu$}
\label{sec:SMEFT}

In this section we present the unitarity bounds for the new physics 
interpretation of 
the muon $g-2$ anomaly 
within a SMEFT approach.   
The strategy consists in fixing the Wilson coefficients 
($C_i / \Lambda^2$)
in terms of the observable $\Delta a_\mu$ 
and determine next the energy scale $\sqrt{s}$ that saturates the unitarity bounds 
derived from the tree-level scattering amplitudes sourced by the effective operator. 
The shorthand $1/\Lambda_i^2 \equiv C_i / \Lambda^2$ is understood in the following.

\subsection{SMEFT approach to $\Delta a_\mu$}

Assuming a short-distance new physics origin of 
$\Delta a_\mu$, 
the leading SMEFT operators contributing up to one-loop order are 
(see Refs.~\cite{Buttazzo:2020eyl,Aebischer:2021uvt} for a more systematic discussion)
\begin{align} 
\label{eq:LSMEFTgm2}
\mathscr{L}_{g-2}^{\rm SMEFT} &= 
\frac{C^\ell_{eB}}{\Lambda^2} (\bar \ell_L \sigma^{\mu\nu} e_R) H B_{\mu\nu} 
+ \frac{C^\ell_{eW}}{\Lambda^2} (\bar \ell_L \sigma^{\mu\nu} e_R) \tau^I H W^I_{\mu\nu} \nonumber \\
&+ \frac{C^{\ell q}_{T}}{\Lambda^2} (\bar \ell^a_L \sigma_{\mu\nu} e_R) 
\varepsilon_{ab} (\bar Q^b_L \sigma^{\mu\nu} u_R) + \text{h.c.} \, , 
\end{align}
which results in \cite{Buttazzo:2020eyl} 
\beq 
\label{eq:gm2fromSMEFT}
\Delta a_{\ell} \simeq \frac{4 m_\ell v}{e\sqrt{2}\Lambda^2} 
\( \Re C^\ell_{e\gamma} - \frac{3\alpha}{2\pi} \frac{c^2_W-s^2_W}{s_W c_W} 
\Re C^{\ell}_{eZ} \log\frac{\Lambda}{m_Z} \) 
- \sum_{q=t,c,u} \frac{4m_\ell m_q}{\pi^2} \frac{\Re C^{\ell q}_T}{\Lambda^2} 
\log\frac{\Lambda}{m_q} \, , 
\eeq
where $C_{e\gamma} = c_W C_{eB} - s_W C_{eW}$ and 
$C_{eZ} = -s_W C_{eB} - c_W C_{eW}$, 
in terms of the weak mixing angle. 
For the Wilson coefficients of the dipole operators 
that contribute at tree level to $\Delta a_{\ell}$, one can 
consistently 
include one-loop 
running,  
obtaining \cite{Degrassi:1998es,Alonso:2013hga}
\beq 
C^{\ell}_{e\gamma} (m_\ell) \simeq C^{\ell}_{e\gamma} (\Lambda) 
\( 1 - \frac{3y^2_t}{16\pi^2} \log\frac{\Lambda}{m_t} - \frac{4\alpha}{\pi} \log\frac{\Lambda}{m_\ell} \) \, . 
\eeq
A convenient numerical parameterization reads 
%\beq 
%\label{eq:DeltaamuSMEFT}
%\Delta a_{\mu} \simeq 
%2.5 \times 10^{-9} 
%\( \frac{270 \ \text{TeV}}{\Lambda} \)^2 
%\(C^\mu_{e\gamma} (\Lambda) - 0.22 \, C^{\mu t}_T (\Lambda)
%- 0.037 \, C^\mu_{eZ} (\Lambda) \) \, , 
%\eeq
\beq
\label{eq:DeltaamuSMEFT}
\Delta a_{\mu} \simeq 
2.5 \times 10^{-9} 
\( \frac{277 \ \text{TeV}}{\Lambda} \)^2 
\( \Re C^\mu_{e\gamma} (\Lambda) - 0.28 \, \Re C^{\mu t}_T (\Lambda)
- 0.047 \, \Re  C^\mu_{eZ} (\Lambda) \) \, , 
\eeq
where we have kept only the leading top-quark contribution for $C_T$ 
(since we are interested on scenarios which maximize the scale of new physics)
and 
the logs have been evaluated for $\Lambda = 277$ TeV.   
Note, however, that the full log dependence will be retained in the numerical analysis below. 
In the following, we will drop the scale 
dependence of the Wilson coefficients, 
which are understood to be evaluated at the scale $\Lambda$. 

\subsection{Unitarity bounds}

Given \eq{eq:LSMEFTgm2}, we can compute the scale of unitarity violation $\Lambda_U$ 
(defined via \eq{eq:LambdaU})
associated with each of the dimension-6 operators involved. 
To do so, we consider here only $2\to 2$ scattering processes,  
since the $2 \to 3$ processes 
(mediated by $\mathcal{O}_{eW}$) turn out to be suppressed by the weak gauge coupling and the 
3-particle phase space, as shown in Appendix \ref{app:2to3scatterings}.
The results obtained by switching one operator per time are collected in Table \ref{tab:SMEFTunitaritybounds}, where 
the bound in correspondence of different initial and final states 
($i \neq f$)
comes from the diagonalization of the scattering matrix (cf.~discussion below \eq{eq:partialwavebound}). 
In \app{app:unitaritySMEFT} we present the full calculation of the unitarity bounds 
stemming from the $\SU(2)_L$ dipole operator, which presents several non-trivial aspects, 
like the presence of higher than $J=0$ partial waves, the multiplicity in $\SU(2)_L$ space and the possibility of 
$2 \to 3$ scatterings. 
\begin{table}[!t]
	\centering
	\begin{tabular}{|c|c|c|c|}
	\rowcolor{CGray} 
	\hline
		Operator & $\Lambda_U$ & $i \to f$ Channels & $J$ \\ \hline
		$\frac{1}{\Lambda^2_{eB}}(\bar \ell_L \sigma^{\mu\nu} e_R) H B_{\mu\nu} $ & $2\sqrt{\pi}|\Lambda_{eB}|$ & $B e_R \to H^\dagger \ell_L$ & $1/2$ \\
		$\frac{1}{\Lambda^2_{eW}}(\bar \ell_L \sigma^{\mu\nu} e_R) \tau^I H W^I_{\mu\nu}$ & $2\sqrt{\pi}\left(\frac{2}{3}\right)^{1/4}|\Lambda_{eW}|$ & $W \bar\ell_L \to H^\dagger \bar e_R$ & $1/2$ \\ 
		$\frac{1}{\Lambda^2_{T,\ell}}(\bar \ell^a_L \sigma_{\mu\nu} e_R) 
\varepsilon_{ab} (\bar Q^b_L \sigma^{\mu\nu} u_R)$ & $2\sqrt{\frac{\pi}{3\sqrt{2}}}|\Lambda_{T,\ell}|$ & $e_R u_R \to Q_L \ell_L $ & $0$ \\
\hline
	\end{tabular}	
	\caption{\label{tab:SMEFTunitaritybounds} 
	Unitarity violation scale for the SMEFT operators 
	%of \eq{eq:LSMEFTgm2} 
	contributing 
	to $\Delta a_\mu$.
	%The scales $\Lambda_i$ are a shorthand notation 
	%for $\Lambda / \sqrt{C_i}$. 
	}
\end{table}

We next make contact with the physical observable $\Delta a_\mu$, 
whose dependence from the Wilson coefficients can be read off \eq{eq:DeltaamuSMEFT}.  
Turning on one operator per time, we find the following numerical values for the 
unitarity violation scales:  
\begin{itemize}
\item $\mathcal{O}^{\mu}_{eB} \equiv (\bar \ell_L \sigma^{\mu\nu} e_R) H B_{\mu\nu}$
%\beq 
%\Lambda_U = 270 \ \text{TeV} \ \sqrt{4\pi} \sqrt{c_W} \simeq 910 \ \text{TeV} \, .
%\eeq
%\beq \mf{
%\Lambda_U = 277 \ \text{TeV} \ \sqrt{4\pi} \sqrt{c_W} \simeq 920 \ \text{TeV} \, .}
%\eeq
\beq
\label{eq:LamUOB}
\Lambda_U  \simeq 277 \ \text{TeV} \ 2\sqrt{\pi} \sqrt{c_W+ 0.047s_W} \simeq 930 \ \text{TeV} \, .
\eeq
\item $\mathcal{O}^{\mu}_{eW} \equiv (\bar \ell_L \sigma^{\mu\nu} e_R) \tau^I H W^I_{\mu\nu}$
%\beq 
%\Lambda_U = 270 \ \text{TeV} \ \sqrt{4\pi} \left(\frac23\right)^{1/4}\sqrt{s_W} \simeq 600 \ \text{TeV} \, .
%\eeq
%\beq \mf{
%\Lambda_U = 277 \ \text{TeV} \ \sqrt{4\pi} \left(\frac23\right)^{1/4}\sqrt{s_W} \simeq 610 \ \text{TeV} \, .}
%\eeq
\beq 
\label{eq:LamUOW}
\Lambda_U \simeq 277 \ \text{TeV} \ 2\sqrt{\pi} \left(\frac23\right)^{1/4}\sqrt{s_W - 0.047c_W} \simeq 590 \ \text{TeV} \, .
\eeq
\item $\mathcal{O}^{\mu t}_{T} \equiv (\bar \ell^a_L \sigma_{\mu\nu} e_R) 
\varepsilon_{ab} (\bar Q^b_L \sigma^{\mu\nu} u_R)$
%\beq 
%\Lambda_U = 270 \ \text{TeV} \ 2 \sqrt{\frac{\pi}{3\sqrt{2}}} \sqrt{0.22} \simeq 220 \ \text{TeV} \, .
%\eeq
\beq
\label{eq:LamUOT}
\Lambda_U \simeq 277 \ \text{TeV} \ 2 \sqrt{\frac{\pi}{3\sqrt{2}}} \sqrt{0.28} \simeq 240 \ \text{TeV} \, .
\eeq
\end{itemize}
Hence, the scale of new physics is maximized if the origin of $\Delta a_\mu$ 
stems from a dipole operator oriented in the $\U(1)_Y$ direction. 

\begin{figure}[!ht]
\centering
\includegraphics[width=0.48\textwidth]{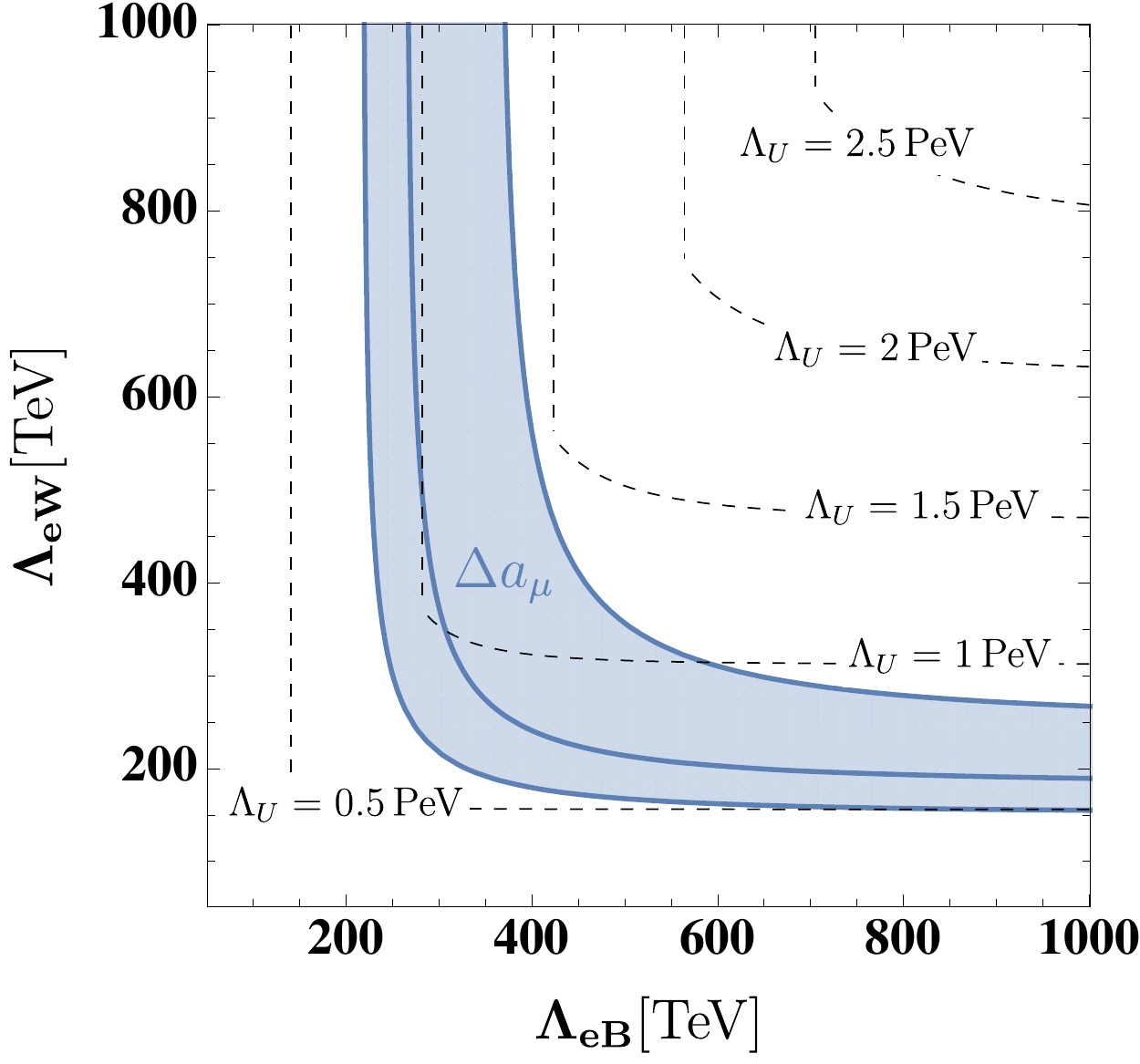}
\caption{In blue, the region in the ($\Lambda_{eB}$, $\Lambda_{eW}$) plane that is 
needed 
to reproduce the experimental value of $\Delta a_\mu$ at the $2\sigma$ level (with the central line corresponding to the central value of $\Delta a_\mu$). The dashed iso-lines represent 
the unitarity bound $\Lambda_U$, defined according to Eq.~\eqref{eq:LU_eBeW}. }
\label{fig:Leb_LeW}       
\end{figure}

If more than one operator is switched on, correlations can arise between the Wilson coefficients whenever they couple same sectors of the theory. For instance, in the 
case in which both the dipole operators 
$\mathcal{O}^\mu_{eW}$ and $\mathcal{O}^\mu_{eB}$ are present 
one can derive a combined bound (see \eq{eq:LU_eBeW}) 
which 
leads to the region displayed in \fig{fig:Leb_LeW}. 
Note that for $\Lambda_{eB} \to \infty$ ($\Lambda_{eW} \to \infty$) 
we reproduce the 
bound with $\mathcal{O}^\mu_{eW}$ ($\mathcal{O}^\mu_{eB}$) only.   
However, if both operators 
contribute sizeably to $\Delta a_\mu$, the unitarity bound can be slightly relaxed above 
the PeV scale.

\section{Renormalizable models}
\label{sec:simplmodels}

We next consider simplified models 
featuring new 
heavy states, which after being integrated out match onto the 
dipole 
and tensor 
SMEFT operators 
contributing to $\Delta a_\mu$
(cf.~\eq{eq:DeltaamuSMEFT}). 
We will then use unitarity methods to set perturbativity limits on renormalizable couplings 
and in turn set an upper bound on the mass of the new on-shell physics states. 
To maximize the scale of new physics,  
we will focus on two renormalizable setups
based scalar-fermion Yukawa theories, allowing for 
a left-right chirality flip that is either entirely due to new physics 
(\sect{sec:oneloopdipole}) or with a top Yukawa insertion (\sect{sec:treetensor}).

\subsection{One-loop matching onto the dipole operator}
\label{sec:oneloopdipole} 
  
\begin{figure}[t!]
\centering
\includegraphics[width=0.9\textwidth]{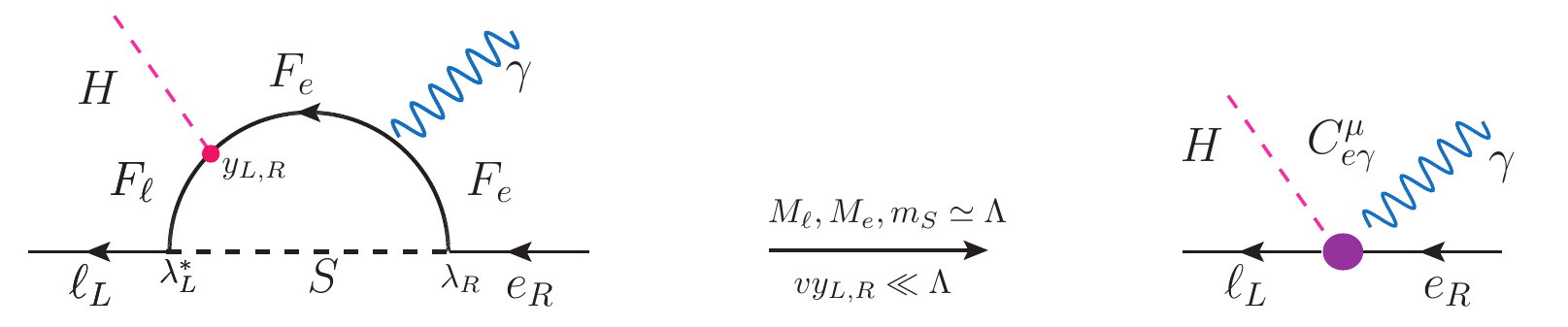}
\caption{Sample diagram of the FFS 
model matching onto $C^\mu_{e\gamma}$ at the scale $\Lambda$.}
\label{fig:Ceg_macthing}       
\end{figure}

In order to match onto the dipole operator at one loop
we consider a simplified model 
with a new complex scalar $S =(1,1,Y+1)$ 
and two vector-like fermions $F_\ell=(1, 2, Y + \frac{1}{2})$ and 
$F_e = (1, 1,Y)$ allowing for a mixing via the SM Higgs (see e.g.~\cite{Calibbi:2018rzv,Arnan:2019uhr,Capdevilla:2020qel,Capdevilla:2021rwo})  
%\begin{equation}
\begin{align}
\label{eq:SFF}
\mathscr{L}^{g-2}_{\rm FFS} &= 
\lambda_L \bar  F_\ell \ell_L S   + \lambda_R \bar  F_e e_R S  + 
\bar F_\ell ( y_{L} P_L + y_{R} P_R  ) F_e H + \rm{h.c.} \nonumber \\
&-  M_\ell \bar F_\ell  F_\ell -  M_e \bar F_e  F_e  - m_ S ^2  |S|^2 
- \kappa\, | H |^2\,  |S|^2 
- \lambda_S |S|^4 \, . 
\end{align}
%\end{equation} 
The FFS model allows for a chirality flip 
of the external leptons 
via the product of couplings $\lambda^*_L y_{L,R} \lambda_R$ 
(cf.~\fig{fig:Ceg_macthing}), 
which can be used to maximize the scale of new physics. 
For $v y_{L,R}\ll M_\ell,M_e,m_S$, we can integrate out the new physics states and find 
at one loop 
\begin{align}
\label{eq:amu5_gen}
\dfrac{C^\mu_{e\gamma}}{\Lambda^2}& 
=-\dfrac{e \lambda_L^* \lambda_R}{32\pi^2 m_{S}^2}\dfrac{\sqrt{x_\ell x_e}}{(x_\ell - x_e)}\,
\Bigg\{
Q_{S} 
\left[
y_R  \(g_S(x_\ell)-g_S(x_e)\)
+
y_L \(\sqrt{\dfrac{x_\ell}{x_e}} g_S(x_\ell)-\sqrt{\dfrac{x_e}{x_\ell}} g_S(x_e)\)
\right]
\nonumber
\\
&
+Q_{F} 
\left[
y_R  \(g_F(x_\ell)-g_F(x_e)\)
+
y_L \(\sqrt{\dfrac{x_\ell}{x_e}} g_F(x_\ell)-\sqrt{\dfrac{x_e}{x_\ell}} g_F(x_e)\)
\right]
\Bigg\} \,,
%\nonumber 
\end{align} 
where $Q_{S} = Y +1$, 
$Q_{F} = Y$, 
$x_{\ell,e}=M^2_{\ell,e}/m^2_S$
and the loop functions 
are given by 
\beq
\label{eq:FG79}
g_F(x)=\dfrac{x^2  - 4x + 3+ 2 \log x}{2 (x-1)^3}\,, \qquad
g_S(x)=\frac{x^2-2 x\log x -1}{2 (x-1)^3}\,.
\eeq
This result agrees with Ref.~\cite{Crivellin:2021rbq} 
in which the special case $y_{L}=y_{R}$ was considered. 
Note that in \eq{eq:amu5_gen} we already matched onto the photon 
dipole operator at the scale $\Lambda$, 
while the connection with the 
low-energy observable $\Delta a_\mu$ is given in 
\eq{eq:DeltaamuSMEFT}. 

Our goal is to maximize the mass of the lightest new physics state for a fixed value of the Wilson coefficient.  
This is achieved in the degenerate limit 
$m_S = M_\ell = M_e$, 
%$x_\ell = x_e = 1$, 
yielding  
\beq
\label{eq:puta}
\dfrac{C^\mu_{e\gamma}}{\Lambda^2}=-\dfrac{e \lambda_L^* \lambda_R}{384 \pi^2 m_S^2}\[\(1+2 Y\) y_L 
- \(1+4 Y\) y_R \] \,  \simeq 
\dfrac{e Y \lambda_L^* \lambda_R}{192 \pi^2 m_S^2}\(2 y_R- y_L \) \, , 
\eeq
where in the last expression we took $Y  \gg 1$. 

The unitarity bounds for the FFS model are summarized in \Table{tab:unitarityboundsFFS}, 
where in the case of multiple scattering channels the bound corresponds to 
the highest eigenvalue of $a^J$.
We refer to \app{app:unitarityweakly} for further details on their derivation.  
Applying these bounds, the maximum value of the 
combination $|\Re (\lambda_L^* \lambda_R (2 y_R-y_L))|$ 
entering 
\eq{eq:puta} is $\approx 121$, while $|eY| \lesssim 3.5$. Hence, we 
obtain 
\beq 
\Delta a_{\mu} \simeq 
2.5 \times 10^{-9} 
\( \frac{131 \ \text{TeV}}{m_S} \)^2 
\(
\dfrac{e Y}{3.5}\)\(\dfrac{\Re\(\lambda_L^* \lambda_R (2 y_R-y_L )\)}{121}
\)\,,
\label{eq:damuFFS}
\eeq
which shows that the $\Delta a_\mu$ explanation in 
the FFS model requires an upper bound on the mass 
of the new on-shell states of about 130 TeV. 
On the other hand, due to the extra loop suppression, 
it is not possible to saturate the unitarity bound that was obtained within the SMEFT 
(see \eq{eq:LamUOB}).

\begin{table}[th!]
	\centering
\renewcommand{\arraystretch}{1.15}
	\begin{tabular}{|c|c|c|}
	\rowcolor{CGray} 
	\hline
		Unitarity bound & $i\to f$ Channels & $J$ \\ \hline 
				$\left|\Re(\lambda_L^* \lambda_R)\right| < 8\pi$ & $e_R F_{\ell_R} \to e_R F_{\ell_R}$ & $0$ \\
		$\left|\Re(y_L^* y_R)\right| < {8\pi}/{\sqrt{2}}$ & $F_{e_R} \bar F_{e_L} \to F_{e_R} \bar F_{e_L}$ & $0$ \\
		$\left| \Re(y_L^* y_R) \pm \sqrt{4|\lambda_L|^2|\lambda_R|^2 + (y_L^*)^2 y_R^2} \right| < 16\pi$ & $i, f=F_{\ell_R}\bar F_{e_L}, e_R \bar \ell_L$ & $0$\\
$2 |\lambda_L|^2 + |\lambda_R|^2 < 8\pi$ & $i, f= F_{\ell_R} \bar \ell_L, F_{e_L} \bar e_R$ & $0$\\
		$|y_R| < \sqrt{8\pi}$ & $H F_{\ell_L} \to H F_{\ell_L}$ & $1/2$ \\
		$|\lambda_R|^2 + 2|y_L|^2 < 16\pi$ & $i, f =  S e_R, H^\dagger F_{\ell, R}$ & $1/2$ \\
		$\left|\Re(y_L \lambda_L^*)\right| < {8\pi}/{\sqrt{2}}$ & $ i, f= F_{e_R} S^\dagger, e_R H$ & $1/2$ \\
		$|\lambda_R|^2 + \sqrt{32|y_L|^2|\lambda_R|^2 + |\lambda_R|^4} < 32\pi$ & $i, f =S \bar F_{e_L}, H^\dagger \ell_L$ & $1/2$\\
		$\left|\Re(\lambda_L y_L)\right| < {16\pi}/{\sqrt{2}}$ & $\ell_L \bar F_{e_L} \to S H^\dagger$ & $1$ \\
		$\left|\Re(y_L^* \lambda_R)\right| < {16\pi}/{\sqrt{2}}$ & $F_{\ell_R} \bar e_R \to H S$ & $1$ \\
		$|\kappa| < {8\pi}/{\sqrt{2}}$ & $H H^\dagger \to S S^\dagger$ & $0$ \\
%		$Y < \frac{\sqrt{6\pi}}{g_Y} -1\lesssim  12$ & $S B_\mu \to S B_\mu$ & $\frac{1}{2}$\\
		{$|g_Y (Y+1)| < \sqrt{6\pi}$} & $S B_\mu \to S B_\mu$ & $1/2$ \\
\hline
	\end{tabular}	
	\caption{\label{tab:unitarityboundsFFS} 
	Unitarity bounds for the FFS model.  
	}
\renewcommand{\arraystretch}{1.0}
\end{table}

\subsection{Tree-level matching onto the tensor operator}
\label{sec:treetensor}

We now consider a simplified 
model that matches onto the tensor operator $\mathcal{O}^{\mu q}_{T}$. 
The scalar leptoquarks $R_2= (3,2,\tfrac{7}{6})$ and $S_1 = (3,1,-\tfrac{1}{3})$ 
allow for a 
coupling to the top-quark 
with both chiralities (see e.g.~\cite{Dorsner:2016wpm}), 
thus maximizing the effect on $\Delta a_\mu$ via a top-mass insertion. 
Massive vectors 
can also lead to renormalizable extensions, 
but they result at least into a $m_b / m_t$ suppression 
compared to 
scalar extensions 
(see e.g.~\cite{Biggio:2016wyy}).    

\begin{figure}[t]
%\vspace{-0.5cm}
\centering
\includegraphics[width=0.9\textwidth]{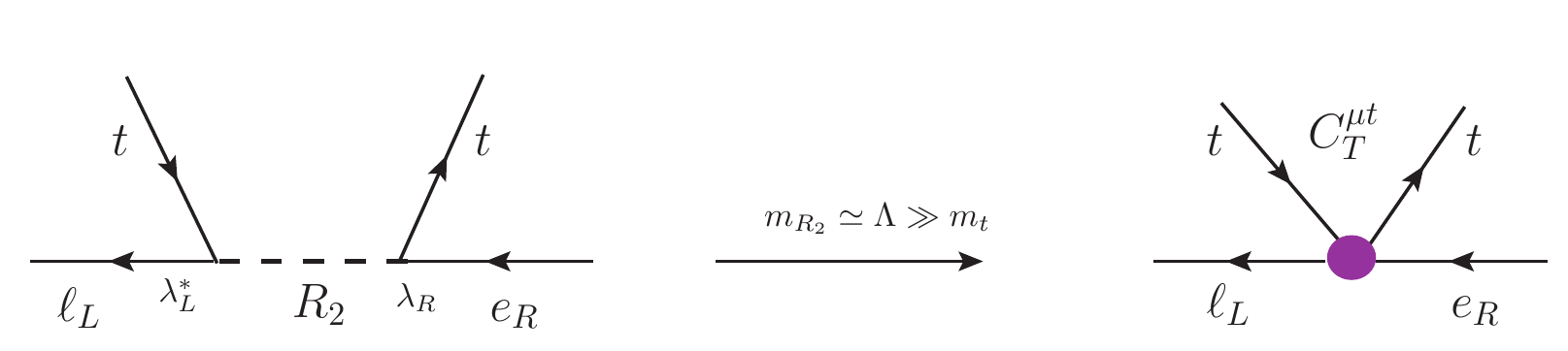}
\caption{Sample diagram of the 
leptoquark
model matching onto $C^{\mu t}_{T}$ at the scale $\Lambda$. }
\label{fig:CT_macthing}       
\end{figure}

Let us focus for definiteness on the $R_2$ case (similar conclusions apply to $S_1$). 
The relevant interaction Lagrangian reads\footnote{Note that 
the leptoquark models in \eq{eq:R2} can be understood as a variant of the FFS model in \eq{eq:SFF}, 
where $F_\ell$ and $F_e$ are replaced by the SM states $q_L$ and $t_R$, 
whereas $S$ is the scalar leptoquark (that is the only new physics state).  
Substituting instead 
$S$ with the SM Higgs and integrating out the heavy $F_\ell$ and $F_e$ fermions 
gives contributions to $\Delta a_\mu$ through dimension-9 SMEFT operators.}
\beq
\label{eq:R2}
\mathscr{L}^{g-2}_{R_2} \supset 
\lambda_L\, \bar t_R \ell_L^a \,\varepsilon_{ab} R_2^b  
+
\lambda_R\, \bar q_L^a\,\mu_R R_{2a} 
+
\text{h.c.} 
\eeq
where $a$ and $b$ are $\SU(2)_L$ indices and $\varepsilon =i \sigma_2$. 
Upon integrating out the leptoquark with mass $m_{R_2} \gg v$ (cf.~\fig{fig:CT_macthing}), 
one obtains~\cite{Feruglio:2018fxo,Aebischer:2021uvt} 
\beq
\dfrac{C^{\mu t}_{T}}{\Lambda^2} =-\dfrac{ \lambda^*_L \lambda_R}{8 m_{R_2}^2 } \, . 
\label{eq:R2matching}
\eeq 
The unitarity bounds for the $R_2$ model (see \app{app:unitarityweakly} for details)
are collected in \Table{tab:unitarityboundsR2} and they imply $|\Re (\lambda^*_L \lambda_R)| \lesssim 12$. Hence, we can recast the contribution to $\Delta a_\mu$ via \eq{eq:DeltaamuSMEFT} 
as 
\beq 
\Delta a_{\mu} \simeq 
2.5 \times 10^{-9} 
\( \frac{180 \ \text{TeV}}{m_{R_2} } \)^2 
\(
\dfrac{\Re(\lambda^*_L \lambda_R)}{12}\)\, .
\label{eq:damuR2}
\eeq
Hence, we conclude that in the leptoquark model one expects 
$m_{R_2}\lesssim 180 \ \text{TeV}$ (the same numerical result is obtained for $S_1$), 
thus providing  
the largest new physics scale 
among the renormalizable extensions 
responsible for $\Delta a_\mu$. 
Moreover, since the matching with the tensor operator is at tree level, 
the leptoquark model fairly reproduces the unitarity bound 
from the SMEFT operator (see \eq{eq:LamUOT}).

\begin{table}[!ht]
\renewcommand{\arraystretch}{1.15}
	\centering
	\begin{tabular}{|c|c|c|}
	\rowcolor{CGray} 
	\hline
		Unitarity bound & $i\to f$ Channels & $J$ \\ \hline
		$|\lambda_L|^2 + |\lambda_R|^2 < 8\pi$ & $i, f=t_R \bar\ell_L, q_L \bar\mu_R$ & $0$\\
		$\left|\Re(\lambda_R\lambda_L^*)\right| < {8\pi}/{\sqrt{3}}$ & $\mu_R\bar\ell_L \to q_L\bar t_R$ & $0$ \\
		$|\lambda_R|^2 < {8\pi}/{3}$ & $q_L R_2^* \to q_L R_2^*$ & $1/2$ \\
		$|\lambda_L|^2 < {16\pi}/{3}$ & $t_R R_2^* \to t_R R_2^*$ & $1/2$ \\
		\hline
	\end{tabular}	
	\caption{\label{tab:unitarityboundsR2} 
	Unitarity bounds for the couplings of the leptoquark model defined in \eq{eq:R2}.}
\renewcommand{\arraystretch}{1.0}
\end{table}

\subsection{Raising the scale of new physics via multiplicity?}

Naively, one could be tempted to increase the upper limit on the scale of new physics by adding 
$\N$ copies of new physics states contributing to $\Delta a_\mu$. 
However, 
while 
both $C_{e\gamma}$ and $C_T$ increase by a factor of $\N$,  
the unitarity bounds on the couplings gets also stronger due to the correlation 
of the scattering channels, so that larger new physics scales cannot be reached. 

In order to see this, consider e.g.~the FFS model with $\N$ copies of $F_\ell$, $F_e$ and $S$.
The scaling of the unitarity bounds is most easily seen in processes where the SM states are exchanged in the $s$-channel, for example $S^i F_{\ell_R}^i \to S^j F_{\ell_R}^j$. Since $\ell_L$ is coupled to all copies in the same way, the $\mathcal{T}$-matrix can be written as
\beq
	\mathcal{T}^{J=1/2} = \frac{1}{32\pi} |\lambda_L|^2 J_\N \,,
\eeq
where $J_\N$ is a $\N \times \N$ matrix filled with 1. Given that the largest eigenvalue of $J_\N$
is $\N$, the unitarity bound on $\lambda_L$ reads
\beq
\label{eq:lamLacsl}
	|\lambda_L| < \sqrt{\frac{16\pi}{\N}} \,.
\eeq 
Similar processes can be considered for all the couplings in Eq.~\eqref{eq:SFF}, leading to a $1/\sqrt{\N}$ scaling for each Yukawa coupling. Hence, the overall $\N$ contribution to 
$\Delta a _\mu \propto 
\N \, \Re (\lambda_L^* \lambda_Ry_{L,R}) / m^2_S$ is compensated 
by the $1 / \sqrt{\N}$ 
scaling of the 
unitarity bounds on the 
couplings and,   
for fixed $\Delta a _\mu$, the mass of extra states gets even lowered at large $\N$. 
In this respect, we reach a different conclusion from the analysis in Ref.~\cite{Capdevilla:2021rwo}.

The same considerations apply if we consider just one new scalar and $\N$ new fermions. The situation is different with $\N$ scalars and just one family of fermions, since $S$ does not couple directly to the Higgs (barring the portal coupling $\kappa$ in \eq{eq:SFF}, which however does not contribute 
to $\Delta a_\mu$). This implies that only $\lambda_L$ and $\lambda_R$ will scale as $1/\sqrt{\N}$, which in turn means that $\Delta a_\mu$ does not change. 
Similar arguments apply when considering larger $\SU(2)_L$ representations, thus implying that the minimal choice we made for the FFS model ensures that $m_S$ is maximized.
The case of the leptoquark $R_2$ is analogous to what we have just described for $\N$ new scalars, with the new fermions of the FFS model replaced by SM fields. Given that $\lambda_L$ and 
$\lambda_R$ would scale as $1/\sqrt{\N}$, there is no gain in taking $\N$ 
copies of leptoquarks.

\section{Non-renormalizable models}
\label{sec:strongmodels}

Till now we focused on renormalizable extensions of the SM 
addressing $\Delta a_\mu$
and showed that they predict on-shell new physics states well below the 
unitarity bound obtained from the SMEFT dipole operators, 
suggesting instead that new physics can hide up to the PeV scale. 
Nonetheless, the SMEFT bound should be understood as the most conservative 
one and applies if the origin of $\Delta a_\mu$ can be 
for instance 
traced back to a 
strongly-coupled dynamics. 
While such a scenario could have calculability issues, we want to 
provide here an intermediate step in which the SMEFT dipole operators 
are generated via a tree-level exchange of a new vector 
resonance from 
a strongly-coupled sector taking inspiration from the case of the $\rho$ meson 
in QCD, but whose UV origin we leave 
unspecified.  

\begin{figure}[t]
%\vspace{-0.5cm}
\centering
\includegraphics[width=0.9\textwidth]{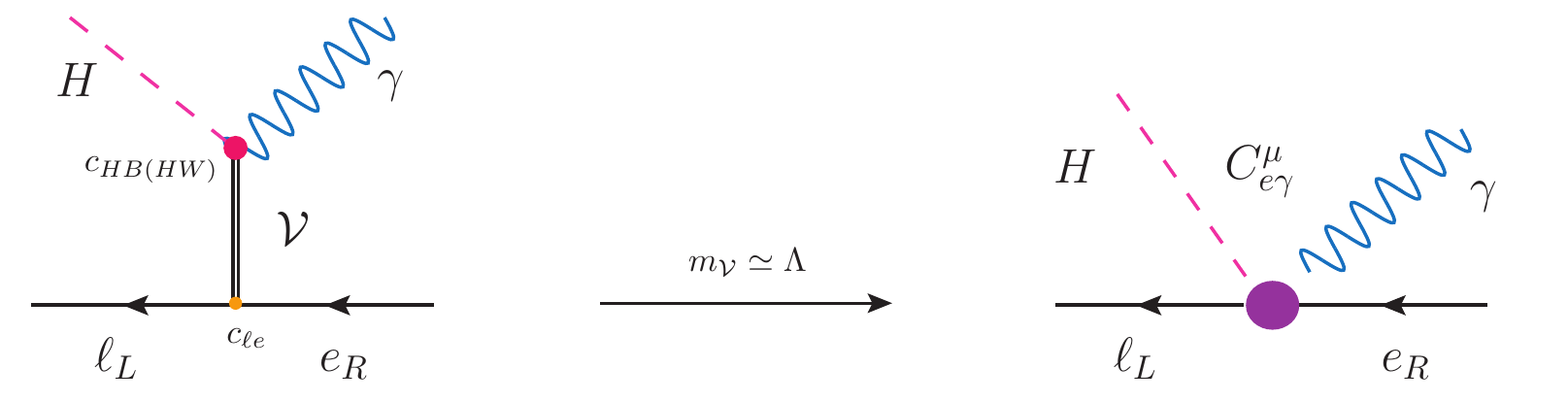}
\caption{Tree-level matching 
onto 
the photon dipole operator 
via the exchange of 
a spin-1 vector resonance.}
\label{fig:KRmatch}       
\end{figure}

Spin-1 vector resonances are conveniently described via 
the two-index anti-symmetric tensor field 
$\B_{\mu\nu}$, following the formalism of Ref.~\cite{Ecker:1988te}. 
In particular, we consider a composite spin-1 state 
featuring the same gauge quantum numbers of the SM Higgs doublet  
and described via the 
effective Lagrangian 
\begin{align}
\label{eq:KRLag}
\mathcal{L}_{\B} &= 
- \D^\mu \B_{\mu\nu}^\dagger \D_\rho \B^{\rho\nu} 
+ \dfrac{1}{2} m^2_{\B} \B_{\mu\nu}^\dagger \B^{\mu\nu} \nonumber \\ 
& +  c_{HB} \B_{\mu\nu}^\dagger  H B^{\mu\nu} + {c}_{HW} \B_{\mu\nu}^\dagger  
 \tau^I H W^{I,\mu\nu}    +  c_{\ell e} \B_{\mu\nu}   (\bar \ell_L \sigma^{\mu\nu} e_R) + \ldots \, ,  
\end{align}
where we neglected $\B_{\mu\nu}$ self-interactions as well as other higher-dimensional operators. 
In fact, \eq{eq:KRLag} should be understood as the leading term of an effective non-renormalizable Lagrangian, with cut-off scale $\Lambda_{\B}$ above $m_{\B}$. 
The free Lagrangian of \eq{eq:KRLag} propagates three degrees of freedom 
describing a free spin-1 particle of mass $m_{\B}$, with propagator \cite{Ecker:1988te,Cata:2014fna,Pich:2016lew} 
\begin{align} 
\label{eq:prop} 
i\Delta_{\mu\nu;\rho\sigma}(q) &= 
\frac{2i}{m^2_\B-q^2}\[
{\cal I}_{\mu\nu;\rho\sigma}(q)
-\dfrac{q^2}{m^2_\B}{\cal P}_{\mu\nu;\rho\sigma}(q)
\]
\, , 
\end{align}
where ${\cal I}_{\mu\nu;\rho\sigma}=\(g_{\mu\rho} g_{\nu \sigma}- g_{\mu\sigma} g_{\nu \rho}\) /2$ and 
 ${\cal P}^{\mu\nu;\rho\sigma}=\(P_T^{\mu\rho} P_T^{\nu\sigma}-P_T^{\mu\sigma} P_T^{\nu\rho}\)/2$ with $P_T^{\mu\nu}=g^{\mu\nu}-q^\mu q^\nu/q^2$. 
Assuming that there is a calculable regime where one can parametrically keep 
$m_{\B} \lesssim \Lambda_{\B}$ 
(in analogy to the chiral approach to the $\rho$ meson in QCD, for which 
$m_\rho \lesssim \Lambda_{\chi} \sim 1$ GeV) we 
can integrate $\B_{\mu\nu}$ out and get 
the 
following
tree-level matching contribution with the  
photon dipole operator (cf.~also \fig{fig:KRmatch}) 
\begin{align}
\label{eq:KRLag2}
%\mathcal{L}^{g-2}_{\B} &\supset 
%\frac{2 y_{H} y_{\B}}{m^2_{\B}}  (\bar \ell_L \sigma^{\mu\nu} e_R) H B_{\mu\nu}  \, . 
\frac{C^\mu_{e\gamma}}{\Lambda^2} =  -\frac{2  \(c_W c_{HB}-s_W {c}_{HW}\) c_{\ell e} }{m^2_\B} \, .
%c_W \frac{2 c_{H} c_{\ell}}{m^2_\B} 
\end{align}
Hence, we obtain
\beq 
\label{eq:DamuKR}
\Delta a_{\mu} \simeq 
2.5 \times 10^{-9} 
\( \frac{1 \ \text{PeV}}{m_\B}\)^2 \( \frac{\Re\(\(-c_W c_{HB}+s_W {c}_{HW}\) c_{\ell e}\)}{7.5} \) \, ,  
\eeq
where we normalized $m_\B$ at the PeV scale, that is in the ballpark of the unitarity bound 
obtained from the 
SMEFT dipole operators. 
It should be noted that although the operators in the second line of \eq{eq:KRLag} have canonical 
dimension equal to 4, scattering 
amplitudes involving the $c_{HB,HW,\ell e}$ couplings, 
as e.g.~$H B \to e_R \bar\ell_L$, 
grow like $s / m^2_\B$ due to the 
high-energy behaviour of the propagator in \eq{eq:prop}. 
Hence, the effective description of the vector resonance breaks down not far 
above $m_\B$, being the theory non-renormalizable.\footnote{Another way to generate the dipole operators 
%$C_{eB,W}$
relevant for 
$\Delta a_\mu$ 
at tree level 
is to consider non-renormalizable 
models, involving for example 
a new vector-like fermion $\mathcal{F} = (1,2,-\tfrac{1}{2})$ \cite{deBlas:2017xtg}.}
%\begin{align}
%\mathscr{L}^{g-2}_{F} \supset c_{FH} \bar{F}_{L} H e_{R} 
%  + \frac{1}{M_F}c_{FB} \(\bar{F}_{R}   \sigma^{\mu\nu} l_{L}\) B_{\mu\nu}
%  + \frac{1}{M_F} c_{FW} \(\bar{\ell}_{L}\sigma^{\mu\nu} \tau^I F_{R}\) W^I_{\mu\nu}
%\end{align}
%\begin{align}
%\frac{C^\mu_{e\gamma}}{\Lambda^2}= &
%     \frac{c_{FH}}{M_{F}^2} \(c_W c_{FB}-s_W c_{FW}\),
%%\\
%%C_{eB}= &     \frac{c_{FB}^* c_{FH}}{M_{F}^2},\qquad
%% C_{eW}=     \frac{c_{FW}^*  c_{FH}}{M_{F}^2}
%\end{align}

\section{Conclusions}
\label{sec:conclusions}

Unitarity bounds are a useful tool in order to infer the 
regime of validity of a given physical description. 
In EFT approaches, the energy scale at which unitarity is violated in tree-level 
scattering amplitudes can be often associated to the onset of the new physics 
completing the effective description. Instead, within 
renormalizable setups unitarity bounds are a synonym of 
perturbativity bounds on the size of the adimensional couplings.
In this work we have investigated unitarity constraints on the new physics 
interpretation of the muon $g-2$ anomaly. Assuming a short-distance 
SMEFT origin of the latter, 
we have first computed unitarity bounds considering a set of leading 
(dipole and tensor) 
operators contributing to $\Delta a_\mu$. It turns out that the scale 
of tree-level unitarity violation is maximized in the case of dipole operators 
and reaches the PeV scale 
%(taking the central value of $\Delta a_\mu$) 
when both $\U(1)_Y$ and $\SU(2)_L$ dipoles 
are switched on  
%(stretching up to $\lesssim 1.5$ PeV when the 
%current 2$\sigma$ value of $\Delta a_\mu$ is considered
(cf.~\fig{fig:Leb_LeW}). 
Hence, most conservatively, in order to resolve the new physics origin of the 
SMEFT operators behind $\Delta a_\mu$ one would need to probe 
high-energy scales up to the PeV. 
This most pessimistic scenario, 
clearly outside from the direct reach of next-generation high-energy particle colliders, 
can be understood as a  
\emph{no-lose theorem} for the muon $g-2$ puzzle. 
Of course, the new physics origin 
of $\Delta a_\mu$ might reside well below the PeV scale, as it is indeed suggested 
by simplified models based on renormalizable scalar-Yukawa theories. 
In the latter case we have considered a couple of well-known scenarios 
matching either on the tensor (at tree level) or the dipole (at one loop) operators of the SMEFT analysis.  
In both cases, we have computed unitarity bounds on renormalizable couplings, 
thus allowing the mass of the new on-shell states to be maximized.  
The latter are found to be $M_{\rm on-shell} \lesssim 130$ TeV and $\lesssim 180$ TeV, 
respectively for the dipole and the tensor operators. 
Moreover, we have shown that multiplicity does not help to relax those bound because 
unitarity limits scale as well with the number of species. 

Since the bound obtained within renormalizable models is well below the 
SMEFT bound, it is fair to ask which UV completions could lead to a 
new physics resolution of the muon $g-2$ puzzle hidden at the PeV scale. 
In fact, one could imagine a strongly-coupled dynamics at the PeV scale that is 
%tautologically 
equivalent to writing the SMEFT Lagrangian.  
Here, we have 
provided instead an intermediate step in which the SMEFT dipole operators 
are generated via the tree-level exchange of a new spin-1 vector resonance 
described by a two-index anti-symmetric tensor field 
$\B_{\mu\nu}$
with the same quantum numbers of the SM Higgs and whose origin 
should be traced back 
to the dynamics of a strongly-coupled sector. 
%which takes inspiration from 
%in analogy to the $\rho$ meson 
%dynamics in QCD. 
%Although we have not provided a model realizing 
This effective scenario provides a non-trivial example in which the 
dipole effective operators are generated via 
tree-level matching, thus suggesting that the SMEFT unitarity bound 
can be saturated with new on-shell states hidden at the PeV scale. 
It would be interesting to investigate whether a UV dynamics leading to 
such effective scenario can be explicitly realized.

\section*{Acknowledgments} 
We thank Paride Paradisi and Bartolomeu Fiol for useful discussions. 
FM acknowledges financial support from the State Agency for Research of the Spanish Ministry of Science and Innovation through the “Unit of Excellence Mar\'ia de Maeztu 2020-2023” award to the Institute of Cosmos Sciences (CEX2019-000918-M) and  from PID2019-105614GB-C21 and  2017-SGR-929 grants.  LA acknowledges support from the Swiss National Science Foundation (SNF) under contract 200021-175940.
The work of MF is supported by the project C3b of the DFG-funded Collaborative Research Center TRR 257, ``Particle Physics Phenomenology after the Higgs Discovery''. The work of MN was supported in part by MIUR under contract PRIN 2017L5W2PT, and by the INFN grant ‘SESAMO’.

\appendix

\section{Unitarity bounds in the SMEFT}
\label{app:unitaritySMEFT}

In this Appendix we expand on some aspects of the calculation of unitarity bounds 
in the SMEFT. 
The case of the operator $\mathcal{O}_{eW} =  (\bar \ell_L \sigma^{\mu\nu} e_R) \tau^I H W^I_{\mu\nu}$ 
is analyzed in detail, since it 
offers the possibility of discussing several non-trivial aspects, like 
the multiplicity of the scattering amplitude in $\SU(2)_L$ space, the contribution of 
higher-partial waves and that of $2 \to 3$ scatterings. 
The calculations of the unitarity bounds for $\mathcal{O}_{eB}$ and $\mathcal{O}_{T}$ follow in 
close analogy and are not reported here.

\subsection{$2 \to 2$ scattering}
\label{app:2to2scatterings}

Consider the  $2 \to 2$
scattering $W^I \bar \ell_L^a \to H^{\dagger,b} \bar e_R$ sourced by $\mathcal{O}_{eW}$, 
where we have explicitly written the $\SU(2)_L$ indices ($I=1,2,3$ in the adjoint 
and $a,b = 1,2$ in the fundamental). Taking a $W$ with positive helicity, 
the lowest partial wave is $J=1/2$. The only possible source for a multiplicity of states in this sector is given 
by $\SU(2)_L$, giving a total of $3 \times 2 + 2 = 8$ states, so the $J=1/2$ sector is a $8\times 8$ matrix, with entries given by $(\tau^I)^{ab}$. Ordering the states as $\{W^1 \bar\ell_L^1,W^1 \bar\ell_L^2,W^2 \bar\ell_L^1,W^2 \bar\ell_L^2,W^3 \bar\ell_L^1,W^3 \bar\ell_L^2, H^{\dagger,1} \bar e_R, H^{\dagger,2} \bar e_R\}$, we have 
\begin{align}
	a^{J=1/2}_{fi} = a_{1/2} 
	\begin{pmatrix}
		0 & 0 & 0 & \tau^1 \\
		0 & 0 & 0 & \tau^2 \\
		0 & 0 & 0 & \tau^3 \\
		\tau^1 & \tau^2 & \tau^3 & 0
	\end{pmatrix} \, , 
	\label{eq:SU2multiplicity}
\end{align}
where $a_{1/2} = \frac{\sqrt{2}}{16\pi}\frac{s}{\Lambda^2_{eW}}$ encodes the result of 
\eq{eq:partialwaves} 
(and whose calculation is reported below). 
The largest eigenvalue of this matrix is $a^{J=1/2}_{ii} = \sqrt{3} a_{1/2}$, leading to the bound
\beq
\label{eq:unitboundeW}
	\sqrt{s} < \Lambda_U = 2 \sqrt{\pi}\left(\frac{2}{3}\right)^{1/4} |\Lambda_{eW}| \,.
\eeq
We now report the computation of the amplitude $a_{1/2}$ of \eq{eq:SU2multiplicity}. 
The process is
\beq
	W(p,+) + \bar\ell_L(k) \to H(p') + \bar e_R (k') \,,
\eeq
with $\vec{p}$ chosen along the $\hat z$ direction and the scattering angle $\theta$ the one formed by $\vec{p}$ and $\vec{p}'$, and we have suppressed $\SU(2)_L$ indices.  
The $\mathcal{T}$-matrix element is
\beq
	\mathcal{T}_{fi} =  \frac{1}{\Lambda^2_{eW}}(p_\mu \varepsilon_\nu^{(+)}(\vec{p}) - p_\nu \varepsilon_\mu^{(+)}(\vec{p}))({\bar v}^{(R)}(\vec{k}) \sigma^{\mu\nu} v^{(L)}(\vec{k}')) = 2\sqrt{2} \frac{s}{\Lambda^2_{eW}} \cos \frac{\theta}{2} \,.
\eeq
Since the lowest partial wave is $J = 1/2$, and $\mu_i = \mu_f = 1/2$, we need the $d$-function $d^{1/2}_{1/2,1/2} (\theta) = \cos \frac{\theta}{2}$. Plugging this into \eq{eq:partialwaves} gives
\beq
	a_{1/2} = \frac{1}{32\pi} 2\sqrt{2} \frac{s}{\Lambda^2_{eW}}  \int_{-1}^1 \mathrm{d} \cos \theta \cos^2 \frac{\theta}{2} = \frac{\sqrt{2}}{16\pi} \frac{s}{\Lambda^2_{eW}} \,.
\eeq

\subsection{$2\to 3$ scattering}
\label{app:2to3scatterings}

Here we show how the unitarity bound for the $2\to 3$ scattering is weaker than 
the one obtained for $2\to 2$ processes, in the special case of the operator $\mathcal{O}_{eW}$. 
This is due to the presence of the weak gauge coupling $g_2 \simeq 0.6$, in addition to the phase-space suppression of the 3-particle final state. Extracting the $2\to 3$ partial wave is slightly more involved, since one needs to construct the three-particle states at fixed total $J$, which in the centre-of-mass frame have five degrees of freedom we have to integrate over, instead of the only two polar angles of the two-particle case.
In particular, a convenient set of variables is the one obtained by combining 2 particles together (as it is done e.g.~for semi-leptonic hadron decays, in which one usually considers the lepton pair). Fixing their mass $m_R^2$, and boosting to the frame in which these are back-to back, one can construct a state with fixed $J_R$ (and helicity $\lambda_R$) out of the two-particles, and then combine this with the third to form the eigenstates of the total angular momentum $J$. The explicit expression is given by
\begin{align}
	\nonumber
	\ket{\sqrt{s},m_R^2;JM;\vec{\lambda}} &= N_J^{(3)}(\vec{\lambda}) \sum_{J_R,\lambda_R} \int \mathrm{d}\Omega_1 \mathrm{d}\Omega_R \mathcal{D}^{J^*}_{M,\lambda_R-\lambda_3}(\phi_R,\theta_R,-\phi_R) \\
	&\times \mathcal{D}^{J_R^*}_{\lambda_R,\lambda_1-\lambda_2}(\phi_1,\theta_1,-\phi_1) \ket{\sqrt{s},m_R^2;\theta_R\phi_R;\theta_1\phi_1;\vec{\lambda}} \,,
\end{align}
where $\mathcal{D}_{MM'}(\alpha,\beta,\gamma)$ are Wigner's $\mathcal{D}$-matrices, with $\alpha$, $\beta$, $\gamma$ Euler angles in the $z$-$y$-$z$ convention and 
\beq
	N_J^{(3)}(\vec{\lambda}) = \frac{\sqrt{2J+1}}{4\pi} \left( \sum_{J_R,\lambda_R} \frac{1}{2J_R + 1} \right)^{-1/2}
\eeq
is a normalisation factor, the angles $\theta_1$ and $\phi_1$ are the polar angles of particle 1 in the centre-of-mass of particles 1 and 2,\footnote{The $\hat z$ axis is chosen along the direction of $\vec{p}_1 + \vec{p}_2$ in the 3-particle centre-of-mass.} $\theta_R$ and $\phi_R$ the polar angles of $\vec{p}_1 + \vec{p}_2$ in the centre-of-mass of the three particles (i.e.~$\vec{p}_1 + \vec{p}_2 + \vec{p}_3 = 0$), and $\vec{\lambda}= (\lambda_1,\lambda_2,\lambda_3)$ are the helicities. The dependence on $\vec{\lambda}$ in the normalization factor is implicit, since the helicities determine over which values the sum over $J_R$, $\lambda_R$ runs. This will have to be considered case by case, depending on the type of particles involved and the partial wave one wants to obtain.
With these states, we can extract the $2\to 3$ partial wave at fixed $m_R^2$, in the case of massless particles, as follows:
\begin{align}
	\nonumber
	a^J_{fi} &= \frac{\sqrt{s-m_R^2}}{256\pi^2 \sqrt{s}} \left(\sum_{J_R} \frac{1}{2J_R + 1}\right)^{-1/2} \sum_{J_R} \int \mathrm{d} \cos{\theta_1} \mathrm{d}\cos{\theta_R}  \\
	&\times d^J_{\mu_i,\mu_i}(\theta_R) d^{J_R}_{\mu_i+\lambda_3,\lambda_1-\lambda_2}(\theta_1) \mathcal{M}_{fi}(\sqrt{s},m_R^2;\theta_1,\theta_R;r,s,\vec{\lambda}) \,,
\end{align}
where $r,s$ are the helicities of the incoming particles, and $\mu_i$ is their sum.
The largest eigenvalue is then given by
\beq
	\xi = \sqrt{\int_0^s \mathrm{d} m_R^2 \left[a^J_{fi}(2\to 3;m_R^2)\right]^2} \,.
\eeq
Finally, the full diagonalisation of the $\mathcal T$-matrix is then achieved by considering the multiplicities in helicity and gauge space, which can lead to further enhancements.

In the case of the operator $\mathcal{O}_{eW}$, the largest channel is the $J=1/2$ scattering $He_R \to \ell_L WW$, yielding the bound
\beq
	\sqrt{s} < \Lambda_U^{2\to 3} = \frac{32\pi}{\sqrt{g_2}} \sqrt{\frac{1}{8+\pi^2}\frac{1}{\sqrt{3}}}\,\Lambda_{eW} \,.
\eeq
Comparing this with the $2\to 2$ bound in \eq{eq:unitboundeW},  
$\Lambda_U^{2\to 2} = 2\sqrt{\pi} \Lambda_{eW}$, 
one finds
\beq
	\frac{\Lambda_U^{2\to 2}}{\Lambda_U^{2\to 3}} \simeq 0.3 \sqrt{g_2} \,.
\eeq

\subsection{A combined bound with $\mathcal{O}_{eW}$ and $\mathcal{O}_{eB}$}
\label{app:combined}

Let us examine now the case in which both the operators $\mathcal{O}_{eW}$ and $\mathcal{O}_{eB}$ 
are switched on. 
Consider again the scattering $W \bar \ell_L \to H^\dagger \bar e_R$ mediated by $\mathcal{O}_{eW}$. From the point of view of SM gauge symmetry, the final state forces the process to occur in the (1,2,1/2) representation. The same applies to the process $B \bar \ell_L \to H^\dagger \bar e_R$. We can therefore construct the $\mathcal{T}$-matrix in a similar manner as above. Now ordering the states as $\{W\bar\ell_L, B\bar\ell_L, H^\dagger e_R\}$, we find
\begin{align}
	a^{J=1/2}_{fi} = \Tilde{a}_{1/2}
	\begin{pmatrix}
		0 & 0 & \frac{1}{\Lambda^2_{eW}} A \\
		0 & 0 & \frac{1}{\Lambda^2_{eB}} \mathbf{1}_{2\times 2} \\
		\frac{1}{\Lambda^2_{eW}} A^\dagger & \frac{1}{\Lambda^2_{eB}}  \mathbf{1}_{2\times 2} & 0
	\end{pmatrix}
	\qquad
	A = \begin{pmatrix} \tau^1 \\ \tau^2 \\ \tau^3 \end{pmatrix} \,,
\end{align}
with $\Tilde{a}_{1/2} = \frac{s\sqrt{2}}{16\pi}$. The largest eigenvalue is $a^{J=1/2}_{ii} = \Tilde{a}_{1/2} \sqrt{\frac{3}{\Lambda^4_{eW}} + \frac{1}{\Lambda^4_{eB}}}$, thus we find
\beq\label{eq:LU_eBeW}
	\sqrt{s} < \Lambda_U = \text{min} \left[ 2\sqrt{\sqrt{2}\pi} \left( \frac{3}{\Lambda^4_{eW}} + \frac{1}{\Lambda^4_{eB}} \right)^{-1/4},\ 2 \sqrt{\pi} |\Lambda_{eB}| \right] \,.
\eeq
Hence, as shown in \fig{fig:Leb_LeW}, we can constrain simultaneously 
$C_{eB}$ and $C_{eW}$. It is worth noticing that, following the same procedure with the scattering $B e_R \to H^\dagger \ell_L$ (which minimizes the bound for $\mathcal{O}_{eB}$), i.e.~considering also $W e_R \to H^\dagger \ell_L$, we would still find two independent bounds for the two operators.\footnote{Using this process to give a bound on $\mathcal{O}_{eW}$ alone, we would find, after considering all $\SU(2)_L$ multiplicities, $\Lambda_U = 2\sqrt{\pi}|\Lambda_{eW}|$, which is slightly weaker than the one given in Table \ref{tab:SMEFTunitaritybounds}.} This is due to the fact that the state $W e_R$ transforms as $(1,3,-1)$, which cannot mix into the $\SU(2)_L$ singlet configuration formed by $B e_R$.

\section{Unitarity bounds in renormalizable models}
\label{app:unitarityweakly}

In this section we provide some details about the computation of the unitarity bounds for the simplified models of \sect{sec:simplmodels}. Staring from the case of the $R_2$ leptoquark, whose interactions relevant for the anomalous magnetic moment are described by the lagrangian \eqref{eq:R2},
\beq
\mathscr{L}^{g-2}_{R_2} \supset 
\lambda_L\, \bar t_R \ell_L^a \,\varepsilon_{ab} R_2^b  
+
\lambda_R\, \bar q_L^a\,\mu_R R_{2a} 
+
\text{h.c.} \,,
\eeq
one can see that several $2 \to 2$ scattering processes can be considered, both scalar and fermion mediated. The goal is therefore to analyse all of them, in order to identify which channel gives the strongest bound.
In general, since there is more than one coupling (two in this case), the different channels will yield independent (combined in general) bounds, as in Table \ref{tab:unitarityboundsR2}. The overall bound on the couplings $\lambda_L$ and $\lambda_R$ can then be visualised as the region defined by the intersection of all the individual constraints. In particular, if the interest lies in one specific combination of said couplings, as for example in \eq{eq:R2matching}, one can maximise the function over this region. 

The best way to proceed in order to compute the unitarity bounds is to classify the possible scattering sectors according to their quantum numbers under the SM gauge symmetry, exploiting the fact that different sectors cannot mix due to gauge invariance. As an example, we show here how the bound is obtained when the leptoquark $R_2$ is exchanged in the $s$-channel, i.e.~the gauge quantum numbers are $(3,2,\frac{7}{6})$. The lowest partial wave, giving the strongest bound, is $J=0$ in this case, and the $\mathcal{T}$-matrix takes the form
\beq
	\mathcal{T}^{J=0}_{(3,2,7/6)} = \frac{1}{16\pi} \begin{pmatrix} |\lambda_L|^2 & \lambda_L^* \lambda_R^* \\ \lambda_L \lambda_R & |\lambda_R|^2 \end{pmatrix} \,,
\eeq
where we have ordered the incoming and outgoing states as $\{t_R\bar\ell_L, q_L \bar\mu_R\}$ and we have taken the high-energy limit. Diagonalising,  the unitarity bound for the highest eigenvalue reads
\beq
	|\lambda_L|^2 + |\lambda_R|^2 < 8\pi  \,.
\label{eq:LRcombinedboundR2}
\eeq
All other bounds in Table \ref{tab:unitarityboundsR2} are obtained in a similar way.

The case of the simplified models with one extra scalar and two extra fermions (FFS) is very similar to the case just described, with some complication due to the presence of more fields and couplings, which increases the number of channels one needs to consider. The philosophy, however, is the same: consider all possible processes and identify the strongest independent bounds (the results are in Table \ref{tab:unitarityboundsFFS}). Once this is done, one can extract a bound on the specific combination of the couplings entering the formula for $\Delta a_\mu$.
Finally, the bounds on the parameter $Y$ entering the hypercharges of the fields $F_e$, $F_\ell$ and $S$ have been obtained by considering scattering channels that are completely separated from the ones where the new Yukawa couplings are involved, i.e.~considering initial and final states containing the $B$ boson. This has the twofold advantage of giving an independent bound on $Y$ while also avoiding issues of unphysical singularities arising in the exchange of a massless vector boson.

\begin{small}

\bibliographystyle{utphys.bst}
\bibliography{bibliography}

\end{small}

\end{document}